\documentclass[twocolumn,showpacs,preprintnumbers,amsmath,amssymb]{revtex4}
\usepackage{graphicx}% Include figure files
\usepackage{dcolumn}% Align table columns on decimal point
\usepackage{bm}% bold math

\begin{document}

\preprint{APS/xxx}

\title{Load Distribution on Small-world Networks\\}% Force line breaks with \\
\author{Huijie Yang}
\email{huijieyangn@eyou.com}
\altaffiliation {Corresponding author}
\author{Tao Zhou}%
\author{Wenxu Wang}%
\author{Binghong Wang}%
\affiliation{%
Department of Modern Physics and Nonlinear Science Center,
University of Science and Technology
of China, Hefei Anhui 230026, China\\}%
\author{Fangcui Zhao}
\affiliation{%
College of Life Science and Bioengineering, Beijing University of Technology, Beijing 100022, China\\}%
\date{\today}% It is always \today, today,

\begin{abstract}
 Mapping a complex network to an atomic cluster, the Anderson
localization theory is used to obtain the load distribution on a
complex network. Based upon an intelligence-limited model we
consider the load distribution and the congestion and cascade
failures due to attacks and occasional damages. It is found that
the eigenvector centrality (EC) is an effective measure to find
key nodes for traffic flow processes. The influence of structure
of a WS small-world network is investigated in detail.
\end{abstract}

\maketitle

  Congestion is a fatal problem in many real world complex systems
such as the Internet, the power grid and the transportation
networks, etc. The dynamical mechanism of the occurrence and the
possible ways to alleviate the congestion attract special
attentions in recent years. It is a complex phenomenon that
depends on a large amount of variables, including the dynamical
processes of traffic flows based upon protocols, the topology
structures of the complex networks, the tolerance of the elements
under attacks or occasional damages.

  In the previous works in modeling traffic flows on complex
networks, a basic assumption is that the networks possess regular
(or random) and homogenous structures [1-14]. However, statistical
measurements of structures tell us that the real world networks
deviate from the Erdos-Renyi and regular networks significantly.
The influence of topology on traffic flow becomes the new focus at
present time [15-17]. And the development of complex network
theory makes it possible for us to deal with this problem in
detail.

In the extensive works, models of traffic flows on computer
networks have been studied, where one common feature for all the
models is that the routers route the data packets to their
destinations in a complete intelligent way, that is, a router
finds for a packet the shortest path between the host and the
destination and forward it along this path step by step. These
models can be called intelligence models (IMs). For an actual
network with a large amount of routers, it is hard for all the
routers to synchronize their route tables due to the exponential
increase of source consumption and the dynamics of networks
(adding and/or deleting of nodes and edges). And finding a
shortest path way based upon complete route tables is clearly a
non-trivial task from the perspectives of source consumptions and
algorithms. Recently, new models with limited intelligence, called
intelligence-limited models (ILMs), are proposed to make the
considerations much close to reality [18,19]. This feature of
protocols determines directly the load distribution of the traffic
flows on complex networks and the consequent congestion and
cascade failures.

When the load of traffic flow is light enough, the created packets
can be processed and delivered in time, leading a steady state
being reached after a short transient time. With the increase of
load, congestion may occur for some nodes, which may induce
cascade failures in a complex network. Hence, obtaining the load
distribution at the steady state is one of the basic problems in
understanding the dynamical processes of traffic flow. In this
Letter, mapping the traffic flow to a statistical feature of a
large amount of particles in an atomic cluster, the
molecular-orbitt theory is used to obtain the load distribution at
the steady state on a complex network. This approach can describe
in a unified way the how the network structure and transferring
protocols affect the traffic behaviors.

  Consider a complex network, the routing algorithms of each node
can be illustrated as,

(\ref{eq1}) Each node can generate $\alpha $ packets per time
step.

(\ref{eq2}) Once a packet reaches its destination, it is removed
from the traffic.

(\ref{eq3}) The destinations are distributed homogenously on the
complex network.

(\ref{eq4}) At each time step, the probability for node $i$'s
delivering a packet to node $j$ is $D(i,j)$. It is determined by
the protocols and the structures of complex networks.

Map a complex network to an atomic cluster, the nodes and edges to
atoms and bonds, respectively. The packet flow on a complex
network can be regarded as the statistical features of a large
amount of particles in a large cluster of atoms. Denote the
adjacent matrix of the complex network with $A$, the element
$A_{ij}$ is $1$ and $0$ if the nodes $i$ and $j$ are connected and
disconnected, respectively. The the coupling Hamilton of this
cluster reads,

\begin{equation}
\label{eq1}
\begin{array}{l}
 H = H_0 + V, \\
 {(H_0)} _{ij} = \varepsilon _0 \delta _{ij} , \\
 V_{ij} = A_{ij} \cdot v_{ij} . \\
 \end{array}
\end{equation}
where $\varepsilon_0$ is the site energy of each node and $V_{ij}$
the coupling between the nodes $i$ and $j$.

The probability of a packet's jumping from node $i$ to node $j$
should be,

\begin{equation}
\label{eq2} D(i,j) \propto \left| {V_{ij} } \right|^2.
\end{equation}

The values of $v_{ij} $ can be determined from the transferring
strategies in protocols. As an example we consider the
intelligence-limited model presented in reference [18], named
partial intelligence-limited model (PILM) in this Letter. In that
model the probability for a packet's jumping from node $i$ to node
$j$ is proportional to a power function of the node $j$'s degree,
e.g., $k(j)^{2\alpha }$. The coupling between node $i$ and node
$j$ should be,

\begin{equation}
\label{eq3} V_{ij}^{PILM} \propto A_{ij} \cdot \left[ {k(i) \cdot
k(j)} \right]^\alpha
\end{equation}

Consider a special condition that the route table of each router
contains only the information that whether it is the destination
of a packet or not. Because of this intelligence limitation, the
packets at a node are delivered forward in a random way to the
connected nodes if it is not the destinations. This model is
called complete intelligence-limited model (CILM) in the present
Letter. For this CILM model, the coupling between node $i$ and
node $j$ should be,

\begin{equation}
\label{eq4} V_{ij}^{CILM} \propto A_{ij}
\end{equation}

At each time step, a node $i$ delivers $C(i)$ packets to the nodes
connecting with it. In literature this deliver capacity, $C =
\left\{ {C(i)\left| {i = 1,2,3, \cdots ,N} \right.} \right\}$, is
designed according to the assumption that the packets are
delivered to the destinations along the shortest paths, i.e., the
complete intelligence of routers (IMs). In the ILMs, this capacity
should be re-designed according to the load distribution.

From the coupling Hamilton depicted in $Eq.(\ref{eq1})$, the
Anderson's localization theory [20] can be employed to investigate
the load distribution in complex networks. Consider conditions
with small values of the packet creation rate $\alpha $, steady
states can be reached after a short transient time. For a regular
network the distribution of traffic packets are homogenous on the
whole network. For a general complex network, the periodic
symmetry is broken, which can induce the localization of the
distribution of traffic packets. Denote the eigenvector
corresponding to the maximal eigenvalue $E_{\max } $ of the
coupling matrix $H$ with,

\begin{equation}
\label{eq5} V_{\max } = \left\{ {V_{\max } (i)\left| {i = 1,2,3,
\cdots ,N} \right.} \right\}.
\end{equation}

At a steady state, the load distribution reads,

\begin{equation}
\label{eq6} p_{steady} (i) = \frac{\left| {V_{\max } (i)}
\right|^2}{\sum\limits_{s = 1}^N {\left| {V_{\max } (s)}
\right|^2} },i = 1,2,3, \cdots N.
\end{equation}
This load distribution can reflect directly the structure effects
and protocol effects on the traffic flows.

It should be noted that the packets can aggregate theoretically at
a node without limits. The packets can be regarded as bosons,
which can stay in a same molecular obit simultaneously. In all the
possible molecular orbits, the principal eigenvector is the only
candidate at which the packets can reach their randomly selected
destinations. For the other molecular orbits, there exist some
special nodes with zero values of the occurrence probabilities. In
these states some packets may stay always in a local area and can
not arrive at their destinations. Hence,the packet current will
not reach a steady state. Setting a life time to each packet can
guarantee the number of the packets in the network tends to a
certain constant, but this local-based steady state can not
realize the communications we expected. The steady state under the
random selection of destinations should be the principal
eigenvector.

Hence, the deliver capacity can be designed as,

\begin{equation}
\label{eq7} C = \left\{ {C(i) = (1 + \delta ) \cdot p_{steady} (i)
\cdot \alpha _0 N\left| {i = 1,2,3, \cdots ,N} \right.} \right\},
\end{equation}
where $\delta > 0$ is the redundant capacity designed for each
router, $\alpha _0 $ the maximal value of the designed packet
creation rate.

With the increase of $\alpha $, congestion becomes possible due to
the increase of loads for all the nodes. Theoretically, we can
obtain the critical point as $\alpha _{critical} = (1 + \delta
)\alpha _0 $, at which the load of each node reaches its designed
deliver capacity. But before this critical point occasional
congestions may occur randomly on the network due to the
fluctuation of the loads. As one of the interesting measurements
we consider the re-distribution of the loads and the possible
cascade failures due to occasional congestions. At each time step,
reckon the number of the nodes whose loads overcome their deliver
capacities as the number of new congestions.

The nodes with heavy loads should be key nodes in considering the
traffic flow on a complex network. Hence, the eigenvector
centrality measure [21,22] is suitable to identify key nodes in a
complex network. By this way we can investigate the attack
effects.

Consider the CILM model on WS small-world networks with the
rewiring probability $p_r \in [0,0.2]$. The WS small-world
networks are generated using the model proposed in Ref. [23]. The
size of a network is $N = 3000$ and the right-handed number of
nodes joined with each node is $k = 2$.

Fig.1 shows that the probability distribution function of load
obeys a power-law, the exponent approximates to $\sim 2.16)$.
There exist some nodes with heavy loads, which should be key nodes
for traffic flows in these complex networks.

Fig.2 presents the loads of all the nodes in a WS small-world
network with $p_r = 0.12$, from which we can find that the nodes
labelled $160,564,1092,735$ are the most important nodes.

Randomly selected 30 nodes are removed from this network to
simulate the occasional damages. For each time step we obtain the
re-distributed loads. The nodes whose loads overcome the designed
capacities are regarded as new congestions and removed from the
original complex network. Fig.3 presents the possible cascade
failures due to occasional damages. When the redundant capacity
$\delta $ is large enough, these occasional damages cannot spread
all over the network and the traffic flow is free and uncongested.
There is a critical value of $\delta _c^{occ} = 0.43$, under which
the global cascade failures will occur.

Fig.4 gives the possible cascade failures due to attacks. The key
node labelled $160$ with the heaviest load is removed from the
original network. If the redundant capacity is not larger than the
critical value $\delta_c^{attack}=0.54$, this removal will induce
the global cascade failures. This critical value is significantly
larger than $\delta _c^{occ} $.

\begin{figure}
\scalebox{0.6}[0.6]{\includegraphics{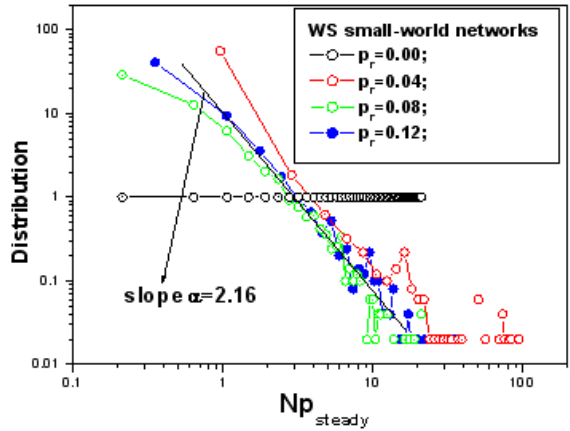}}
\caption{\label{fig:epsart} The probability distribution function
of load for WS small-world networks. These functions obey a
power-law. There exist some nodes with heavy loads, which should
be key nodes for traffic flows in these complex networks.}
\end{figure}

\begin{figure}
\scalebox{0.6}[0.7]{\includegraphics{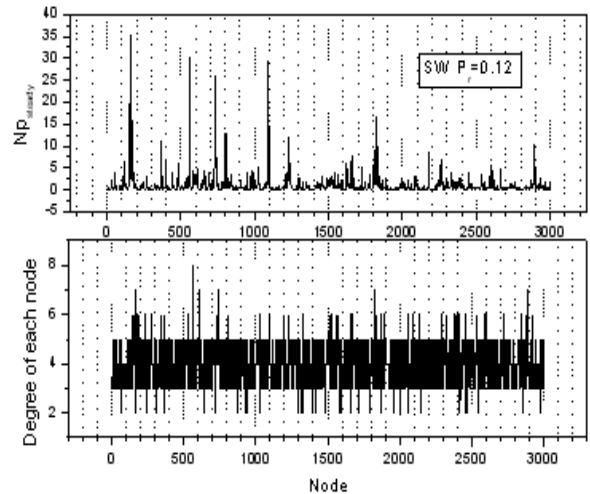}}
\caption{\label{fig:epsart} Load distribution on a WS small-world
complex network. The nodes numbered $160,564,1092,735$ are the
most important nodes.}
\end{figure}

\begin{figure}
\scalebox{0.6}[0.6]{\includegraphics{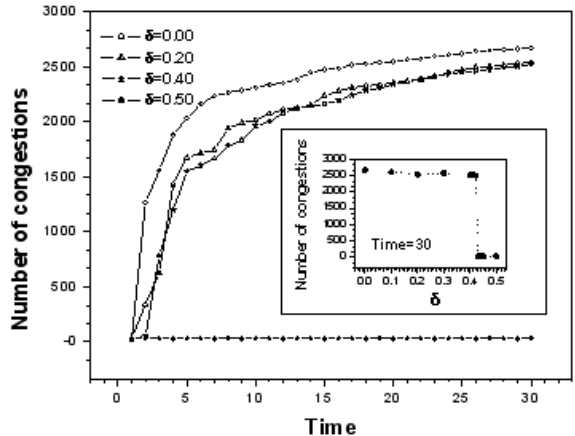}}
\caption{\label{fig:epsart} Cascade failures due to occasional
damages on the WS small-world network($p_r=0.12$). $\alpha_0=10$.
Randomly selected 30 nodes are removed from this complex network
to simulate the occasional damages. For each time step we obtain
the re-distributed loads. The nodes whose loads overcome the
designed capacities are regarded as new congestions and removed
from the original complex network. There is a critical value of
$\delta_{c}^{occ}=0.43$ , under which cascade failures may occur.
In the inset we present the number of congestions at 30 unit time
for different value of redundant $\delta$.}
\end{figure}

\begin{figure}
\scalebox{0.6}[0.6]{\includegraphics{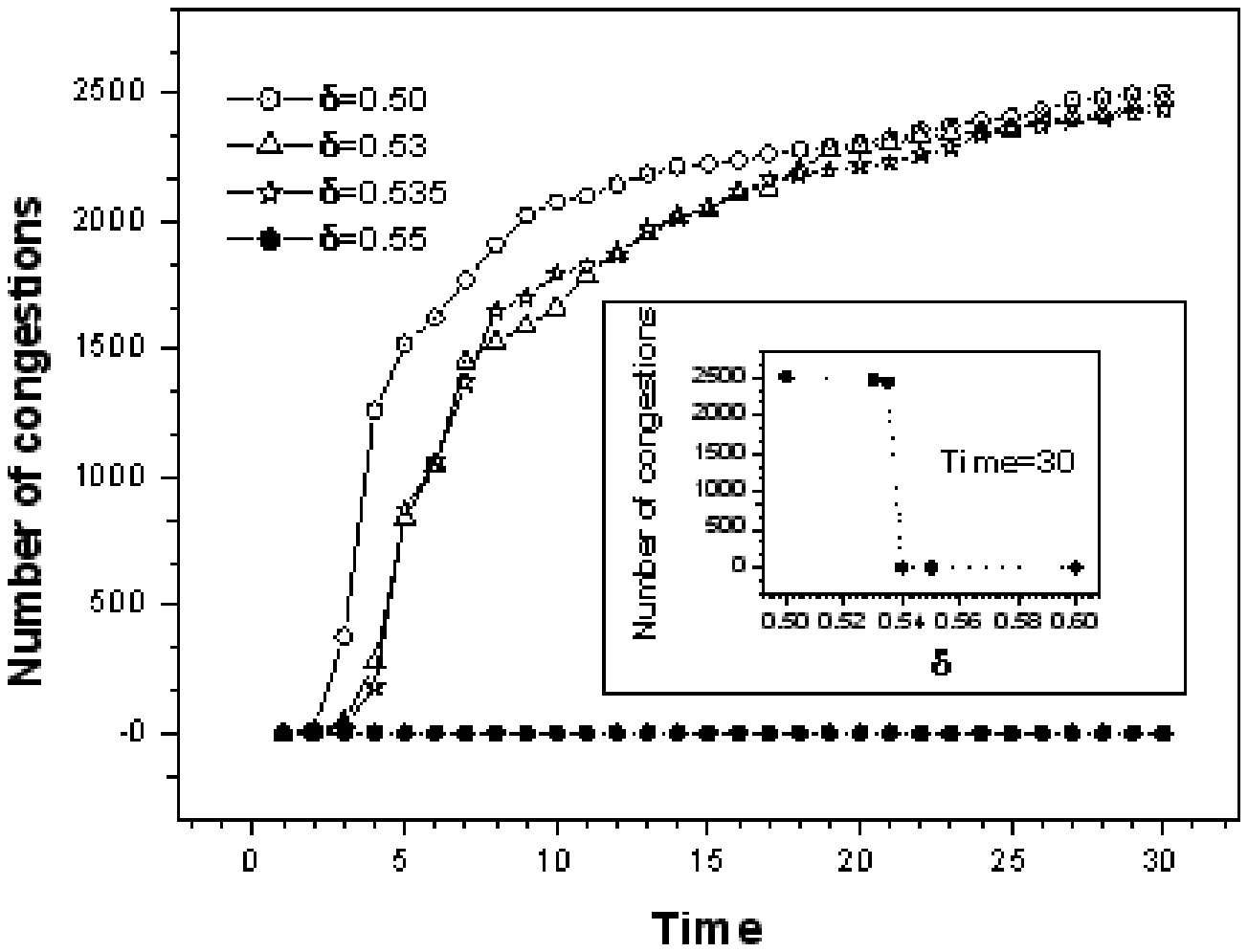}}
\caption{\label{fig:epsart} Cascade failures due to
attacks.$\alpha_0=10$. The key node $No.160$ with the heaviest
load is removed from the original complex network. This removal
can induce cascade failures. The critical value is
$\delta_{c}^{attack}=0.54$ , which is larger than
$\delta_{c}^{occ}=0.43$ significantly. In the inset we present the
number of congestions at 30 unit time for different value of
redundant $\delta$. }
\end{figure}

In summary, mapping the traffic flow of packets to a statistical
feature of a large amount of electrons in an atomic cluster, we
propose the coupling Hamilton to describe the jumping processes
between the nodes of a complex network. The Anderson localization
theory is used to obtain the load distribution on a complex
network at steady state.

This method to determine the load distribution on a complex
network makes it possible to design the delivering capacities of
nodes according to the conditions such as the structure of a
complex network, the packet transferring protocol, the tolerance
performance of each node and even the bandwidths of the edges,
etc. Consequently, from the coupling Hamilton, we can detect the
key nodes in the dynamical process of traffic flow.

The CILM model is used to consider the traffic processes on a WS
small-world network. For this complex network there exist some key
nodes, removal of which may induce cascade failures. Occasional
damages can also induce cascade failures. There are two critical
points, $\delta _c^{occ} $ and $\delta _c^{attack} $, under which
cascade failures can occur due to occasional damages and attacking
the most important key node, respectively. These simulations show
that the method proposed in this paper is powerful and universal
to determine the load distribution on a complex network at steady
state and to detect the key nodes in the dynamical process of
traffic flow.

\begin{acknowledgments}
 This work has been partially supported by the National Natural Science
Foundation of China under Grant No. 70471033, 10472116 and
No.70271070. It is also supported by the Specialized Research Fund
for the Doctoral Program of Higher Education (SRFD No.
20020358009). One of the authors (H. Yang) would like to thank
Prof. Y. Zhuo, Prof. J. Gu in China Institute of Atomic Energy and
Prof. S. Yan in Beijing Normal University for stimulating
discussions.
\end{acknowledgments}

\end{document}